\RequirePackage{fix-cm}
\documentclass[twocolumn]{svjour3}          

\usepackage{geometry}
\usepackage{multirow}
\usepackage{caption}
\usepackage{amssymb}
\usepackage{amsmath}
\usepackage[shortcuts]{extdash}
\usepackage{makecell}
\usepackage{xcolor}
   \definecolor{mygreen}{RGB}{121, 149, 64}
   \definecolor{myblue}{RGB}{0, 112, 192}
   \definecolor{myyellow}{RGB}{191, 144, 0}
   \definecolor{myred}{RGB}{192, 80, 70}

\smartqed  
\usepackage{graphicx}
\usepackage{booktabs}
\usepackage{url}

\begin{document}
\title{Unsupervised domain adaptation based COVID-19 CT infection segmentation network \thanks {This research work is supported by the Air Force Office of Scientific Research (award number FA2386-19-1-4001)}}

\author{Han Chen \and Yifan Jiang \and Murray Loew \and Hanseok Ko}
\institute{Han Chen, \\\email{jessicachan@korea.ac.kr} 
\\ Yifan Jiang, \\\email{yfjiang@korea.ac.kr}
\\ Hanseok Ko (corresponding author), \\\email{hsko@korea.ac.kr}
\at School of Electrical Engineering, Korea University, Seoul 02841, South Korea
\\
\\
Murray Loew \\\email{loew@gwu.edu} \at Department of Biomedical Engineering, George Washington University, Washington D.C., USA
}

\date{Received: date / Accepted: date}
\maketitle

\begin{abstract} 
Automatic segmentation of infection areas in computed tomography (CT) images has proven to be an effective diagnosis approach for COVID-19. However, due to the limited number of pixel-level annotated medical images, accurate segmentation remains a major challenge. In this paper, we propose an unsupervised domain adaptation based segmentation network to improve the segmentation performance of the infection areas in COVID-19 CT images. In particular, we propose to utilize the synthetic data and limited unlabeled real COVID-19 CT images to jointly train the segmentation network. Furthermore, we develop a novel domain adaptation module, which is used to align the two domains and effectively improve segmentation network's generalization capability to the real domain. Besides, we propose an unsupervised adversarial training scheme, which encourages the segmentation network to learn the domain-invariant feature, so that the robust feature can be used for segmentation. Experimental results demonstrate that our method can achieve state-of-the-art segmentation performance on COVID-19 CT images.

\end{abstract}

\section{Introduction}
\label{intro}
The novel coronavirus disease (COVID-19) has become one of the most serious global pandemics. COVID-19 is caused by the infection of severe acute respiratory syndrome coronavirus 2 (SARS-CoV-2), which can be transmitted via breathing, coughing, sneezing, or other means. A recent report \cite{covid19_fatalityrate} showed that, by March 2021, more than 120 million people around the world would have been infected with COVID-19, with a fatality rate of over 2$\%$.

To diagnose COVID-19, the real-time reverse transcription polymerase chain reaction (RT–PCR) test is routinely used. However, RT-PCR is time-consuming, and a series of tests may be required to exclude the possibility of false negatives, which means that there is an urgent need for alternative methods for the fast and accurate diagnosis of COVID-19. Chest computed tomography (CT) has been strongly recommended for the early recognition and evaluation of suspected SARS-CoV-2 infection \cite{xu2020chest}. Chest CT scans are very useful for the auxiliary diagnosis of the typical radiographic features of COVID-19, including ground-glass opacity and consolidation \cite{chung2020ct}. Therefore, the qualitative assessment of infection in CT scans could provide important information in the fight against the spread of COVID-19. Image segmentation has proven to be effective in COVID-19 CT image analysis \cite{gozes2020rapid,shan2020lung,shen2020quantitative}, but it remains challenging because (1) the diversity in the size and distribution of infection leads to a large number of false negative segmentation results, and (2) ground-glass opacity and consolidation are similar in appearance, this small inter-class difference makes the segmentation more difficult \cite{shi2020radiological}. 

Deep learning based automatic segmentation is a powerful technique for medical imaging analysis \cite{shi2020review}. The excellent performance can be attributed to the availability of large volumes of labeled training data. However, it is time-consuming and laborious to collect a sufficient number of COVID-19 CT images with annotations due to concerns over patient privacy \cite{adler2012sharing,sharma2019preserving} and lack of experts \cite{dai2019transfer}. To tackle this issue, some methods have employed parameterized transformation to augment the limited annotated COVID-19 CT images for supervised learning \cite{zheng2020deep,huang2020serial,qi2020machine,zhou2020automatic,elharrouss2020encoder,xu2020gasnet,ouyang2020dual,oulefki2021automatic}. Despite parameterized transformation can solve the problem of data shortage to some extent, but networks trained on limited data still suffer from the poor generalization to unseen datasets due to the insufficient data diversity. Besides, several works have explored to construct new networks suitable for small-scale labeled COVID-19 data \cite{fan2020inf,qiu2020miniseg,laradji2020weakly}, of which the high performance relies on the carefully designed network structure, thus losing scalability and flexibility.

More recently, some efforts have been devoted to generating synthetic COVID-19 CT data for promoting computer-aided diagnosis ability of COVID-19 \cite{liu20203d,jiang2020covid,li2020ct}, which made it possible to train deep models on synthetic images and computer-generated annotations. Nevertheless, as the work \cite{9413443} shows, a model directly trained with the synthetic data may fail to produce precise results for real COVID-19 CT images due to the domain shift. In view of the fact that (1) existing supervised and semi-supervised methods are limited by small-scale COVID-19 CT data; (2) the synthetic COVID-19 CT data is not available directly for training due to domain shift problem. A natural and 
practical question comes up: how to properly utilize the potential of synthetic data to improve the segmentation performance on COVID-19 CT images?

To address above issues, we propose a novel unsupervised domain adaptation based segmentation network for COVID-19 CT infection segmentation task. The contributions of this paper can be summarized as follows: (1) we propose to make full use of synthetic data and limited unlabeled real COVID-19 CT images to jointly train the segmentation network, so as to introduce richer diversity; (2) we design a domain adaptation module to align the two domains and overcome the domain shift. It effectively improves the generalization capability of segmentation network; (3) we propose an unsupervised adversarial training scheme, in which the cross-domain adversarial loss will guide the segmentation network to learn domain-invariant feature, thus improving the segmentation performance. In the meanwhile, our training scheme is very flexible, as it can be arbitrarily combined with any segmentation network with encoder-decoder structure.

The remainder of this paper is organized as follows. In Section \ref{Realted works}, we review previous related works. Section \ref{Methodology} discusses the main components and training scheme of our proposed method, while Section \ref{Experiments} describes our experiments on real COVID-19 CT images. Finally, Section \ref{Conclusion} concludes the paper.

\section{Related works}
\label{Realted works}

\noindent \textbf{COVID-19 infection segmentation.} 
Medical imaging such as CT has played an important role in the fight against COVID-19. As an essential step in the processing and assessment of CT images, segmentation can identify the regions of interest, such as ground-glass opacity, consolidation, and the lung \cite{shi2020review}. Recently, deep learning based segmentation methods have been utilized in COVID-19 CT diagnosis \cite{zheng2020deep,huang2020serial,qi2020machine,zhou2020automatic,elharrouss2020encoder,xu2020gasnet,ouyang2020dual,oulefki2021automatic,fan2020inf,qiu2020miniseg,laradji2020weakly}. For instance, Ouyang et al. \cite{ouyang2020dual} developed a 3D CNN network for COVID-19 infection segmentation, and proposed a dual-sampling attention mechanism to alleviate the imbalanced problem of data. Oulefki et al. \cite{oulefki2021automatic} presented a multilevel thresholding procedure based on Kapur entropy to improve the COVID-19 segmentation performance. Fan et al. \cite{fan2020inf} presented a semi-supervised segmentation method based on random selection propagation strategy, which requires only a few labeled images and primarily utilizes unlabeled data. Qiu et al. \cite{qiu2020miniseg} proposed a lightweight network to solve the overfitting problem caused by the limited training data for COVID-19 segmentation. Laradji et al. \cite{laradji2020weakly} proposed a new COVID-19 segmentation model using point-level rather than full image-level annotations, which overcame the labeling issue to some extent. Most previous works are trained with a supervised or semi-supervised manner, thus the performance is limited by the scale of the labeled data. Furthermore, since the infection areas of COVID-19 could be small with large variations of shapes and textures, segmenting the areas of infection is still a challenging task.

\noindent \textbf{Unsupervised medical segmentation.}
To deal with the lack of annotated data, unsupervised segmentation techniques have attracted growing interest. Most existing proposals employ clustering, which divides a medical image into different groups according to the similarity of the image intensity. For example, Jose et al. \cite{jose2014brain} proposed a method based on K-means clustering to segment abnormal brain regions. Cheng et al. \cite{ouyang2020self} utilized a clustering algorithm to generate superpixel-based pseudo-labels to provide supervision for the segmentation network. However, these clustering based methods heavily depend on the pixel intensity, which means different areas but with similar intensity are likely to be mistakenly segmented into the same class. This is undesired in COVID-19 infection segmentation, because ground-glass opacity and consolidation may not be distinguished as they have similar appearance. Some studies regard the unsupervised medical segmentation task as an unsupervised deformable registration process \cite{shan2017unsupervised,de2019deep,xu2019deepatlas}. Despite the success of these methods, they are insufficient for COVID-19 segmentation since there are large variations of infection on CT images, such as irregular shapes and ambiguous boundaries \cite{wang2020noise}.

\noindent \textbf{Domain adaptation.} Domain adaptation aims to reduce the shift between two distributions \cite{patel2015visual,wang2018deep}, it has been widely employed in conjunction with the use of synthetic data for real-world tasks \cite{lee2020strdan,kim2020learning,shao2020domain,yang2020one,hsu2020progressive}. There are several different strategies proposed to gain better domain adaptation. Some studies utilize maximum mean discrepancy (MMD) \cite{gretton2012kernel} to minimize differences between feature distributions \cite{ghifary2014domain,yan2017mind,haeusser2017associative}, but its effect is limited by whether the distributions follows Gaussian distribution. Another strategy is self-training, which utilizes predictions from an ensemble model as pseudo-labels for unlabeled data to train the current model \cite{li2018disaster,zou2018unsupervised,spadotto2021unsupervised}. There is increasing interest in the use of adversarial training to achieve domain adaptation \cite{vu2019advent,tzeng2017adversarial,pan2020unsupervised,park2020fusion}. This approach reduces the domain shift by forcing the features from different domains to fool the discriminator, thus leading to features from different domains exhibiting a similar distribution. For medical image segmentation, domain adaptation has also demonstrated positive effects \cite{perone2019unsupervised,degel2018domain,zhang2018task,ren2018adversarial,kamnitsas2017unsupervised}. For instance, Degel et al. \cite{degel2018domain} minimized segmentation loss with a domain discriminator to encourage feature domain-invariance across ultrasound datasets for left atrium segmentation. Christian et al. \cite{perone2019unsupervised} addressed the domain shift by extending the self-ensembling method to MRI image segmentation. Kamnitsas et al. \cite{kamnitsas2017unsupervised} employed adversarial learning and utilized synthetic data and sufficient labeled data for brain lesion segmentation. 

The domain adaption technology has achieved some impressive success, especially in the medical imaging field. Therefore, we consider exploiting this novel technology to solve COVID-19 CT infection segmentation task in this paper.

\begin{figure*}
\centering
\includegraphics[width=12cm]{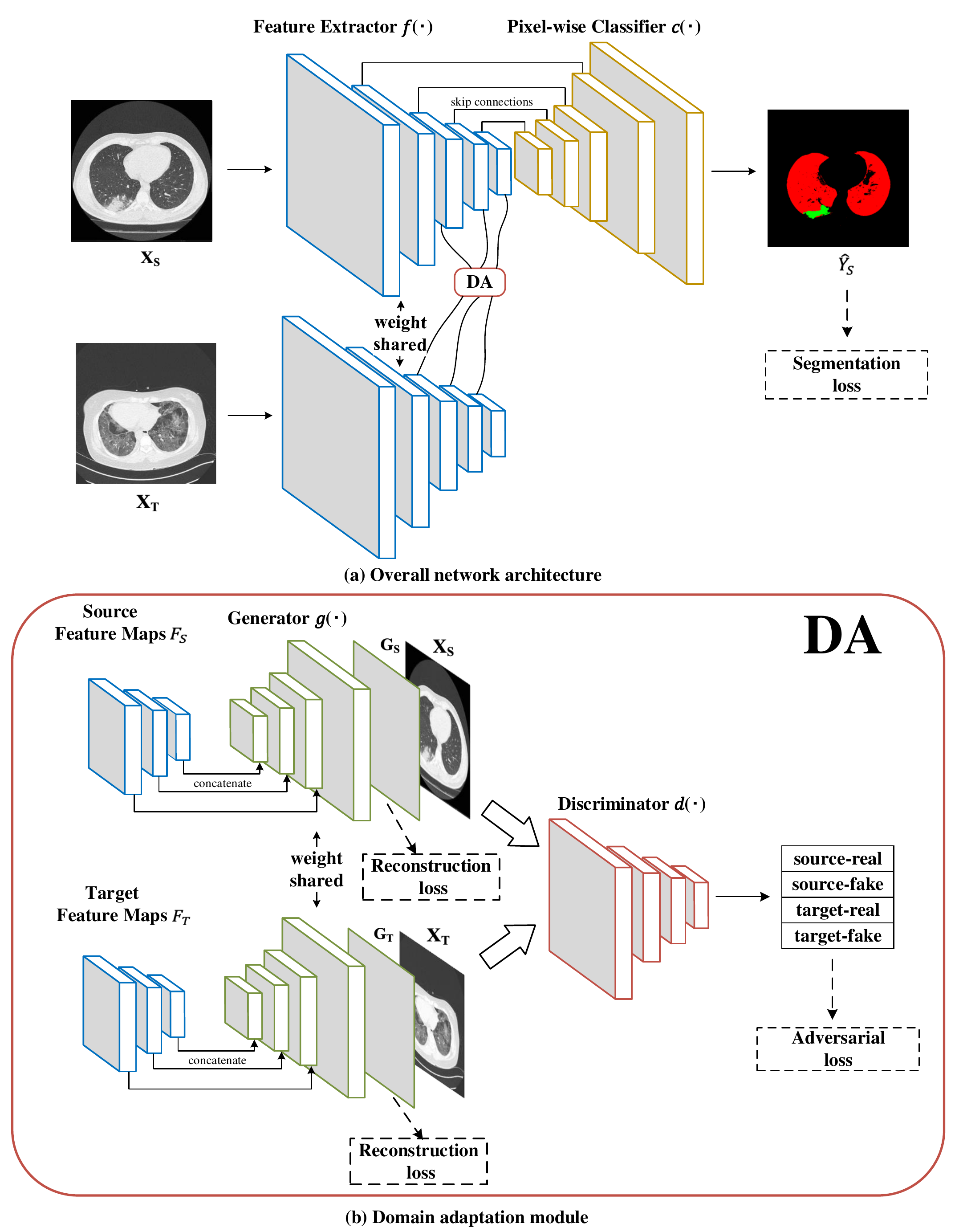}
\caption{Overview of the proposed network. Our network consists of two parts: the segmentation network including a \textcolor{myblue}{feature extractor}, a \textcolor{myyellow}{pixel-wise classifier}, as well as the domain adaptation module (DA) including a \textcolor{mygreen}{generator} and a \textcolor{myred}{discriminator}. The black solid lines with one-way arrow indicate the data flow and the dashed lines denote reconstruction and adversarial loss. The feature extractor and pixel-wise classifier together perform the segmentation task. The DA module is introduced to overcome the domain shift through adversarial training in image space.}
\label{fig:overall}       
\end{figure*}

\section{Methodology}
\label{Methodology}
\subsection{Overview of the proposed method}
As shown in Figure \ref{fig:overall}, our method consists of two parts: the segmentation network composed of a feature extractor ${f(\cdot)}$ and a pixel-wise classifier ${c(\cdot)}$, as well as the domain adaptation module including a generator ${g(\cdot)}$ and a discriminator ${d(\cdot)}$. The source dataset (the synthetic data) and target dataset (the COVID-19 CT images without annotations) are denoted as $\{\mathcal{X}_S, \mathcal{Y}_S\}$ and $\{\mathcal{X}_T\}$, respectively. We first forward the two inputs $X_S$ and $X_T$ to ${f(\cdot)}$, generating feature maps $F_S$ and $F_T$. Then, the ${c(\cdot)}$ takes $F_S$ as input and produces an image-sized segmentation map $\hat{Y}_S$, which is used to optimize segmentation network together with $Y_S$. To overcome the domain shift, we align the distributions of the source and target data using domain adaptation module in the image space. We utilize ${g(\cdot)}$ to reconstruct the inputs conditioned on the feature maps $F_S$ and $F_T$. We then feed the outputs of ${g(\cdot)}$ and $X_S$, $X_T$ to the discriminator ${d(\cdot)}$ and classify them as real or fake within- or cross-domain. The gradients of the cross-domain adversarial loss are propagated from ${d(\cdot)}$ to ${f(\cdot)}$, which leads ${f(\cdot)}$ to learn transferable feature representations applicable to both the source and target domains.

\subsection{Network structure}
\label{Network structure}
\noindent \textbf{Feature extractor.} We build a feature extractor that follows the typical architecture of convolutional neural network. It is composed of four $3\times3$ convolutional layers, and each is followed with a $2\times2$ max pooling operation. Given a source image ${X_S}$ and a target image ${X_T}$, the feature extractor shares the weights and produces feature maps $F_S$ and $F_T$, as shown in equation (\ref{eq1}),

\begin{equation}
\left. F_\delta=f(X_\delta), \delta\in{(S,T)}\right.
\label{eq1}
\end{equation}

\noindent where $\delta\in{(S,T)}$ denotes whether the term stems from the source domain or target domain. The learned features are then sent to the classifier and generator. The former is used to generate pixel-level segmentation results, while the latter is projected into image space for further domain adaptation. 

\noindent \textbf{Pixel-wise classifier.} With the learned feature maps, the pixel-wise classifier converts low-resolution, semantically strong features into pixel-wise classification results, i.e., a class label is assigned to each pixel. We build a classifier that contains three upsampling layers, and each layer is followed by a concatenation with the correspondingly cropped feature map from the feature extractor. It takes $F_S$ as input and produces segmentation map $\hat{Y}_S$ with the same size as ${X_S}$, i.e.,

\begin{equation}
\left. \hat{Y}_S=c(F_S)\right.
\label{eq2}
\end{equation}

As discussed later, in order to make our network have the pixel-level discriminative ability, we use the above predicted segmentation map to calculate the segmentation loss in a supervised manner. Because we can only access the annotations of the source data, we feed only the feature maps of the source domain to the classifier to obtain the segmentation map.

\noindent \textbf{Domain adaptation module.} Unlike recent adversarial based domain adaptation approaches for segmentation tasks that directly calculate the adversarial loss in feature space. Here, we utilize the generator to project the intermediate feature maps to image space for robust adversarial training. Given the feature maps $F_S$ and $F_T$, the generator shares the weights and produces the reconstructions of source image ${X_S}$ and target image ${X_T}$. The reconstructions ${G_S}$, ${G_T}$, and ${X_S}$, ${X_T}$ are then sent to the discriminator and classified as real or fake. The reconstruction process is formulated as equation (\ref{eq3}),

\begin{equation}
\left. G_\delta=g(F_\delta), \delta\in{(S,T)}\right.
\label{eq3}
\end{equation}

Image-space reconstruction is more robust than feature map when applied to the calculation of adversarial loss. This is particularly so for our infection segmentation task, where the differences in the intensity and texture between the source and target images are not that significant. Sub-section \ref{Ablation study} provides detailed verification of the effectiveness of the image-space training. 

The design of our domain adaptation module is inspired by PatchGAN \cite{zhu2017unpaired}. Our generator consists of four upsampling layers, each layer is composed of a $3\times3$ transposed convolutional layer and two residual blocks \cite{he2016deep}. Our discriminator includes two $4\times4$ convolutional layers, and the first layer is followed by nine residual blocks. For each input, the output of domain adaptation module is a probability map, in which the value of each pixel indicates the possibility that each patch in the input is real or fake. Compared with a normal discriminator whose output is only real numbers, our discriminator is more helpful for retaining detailed information. 

\begin{figure}[!t]
\centering
\includegraphics[width=8cm]{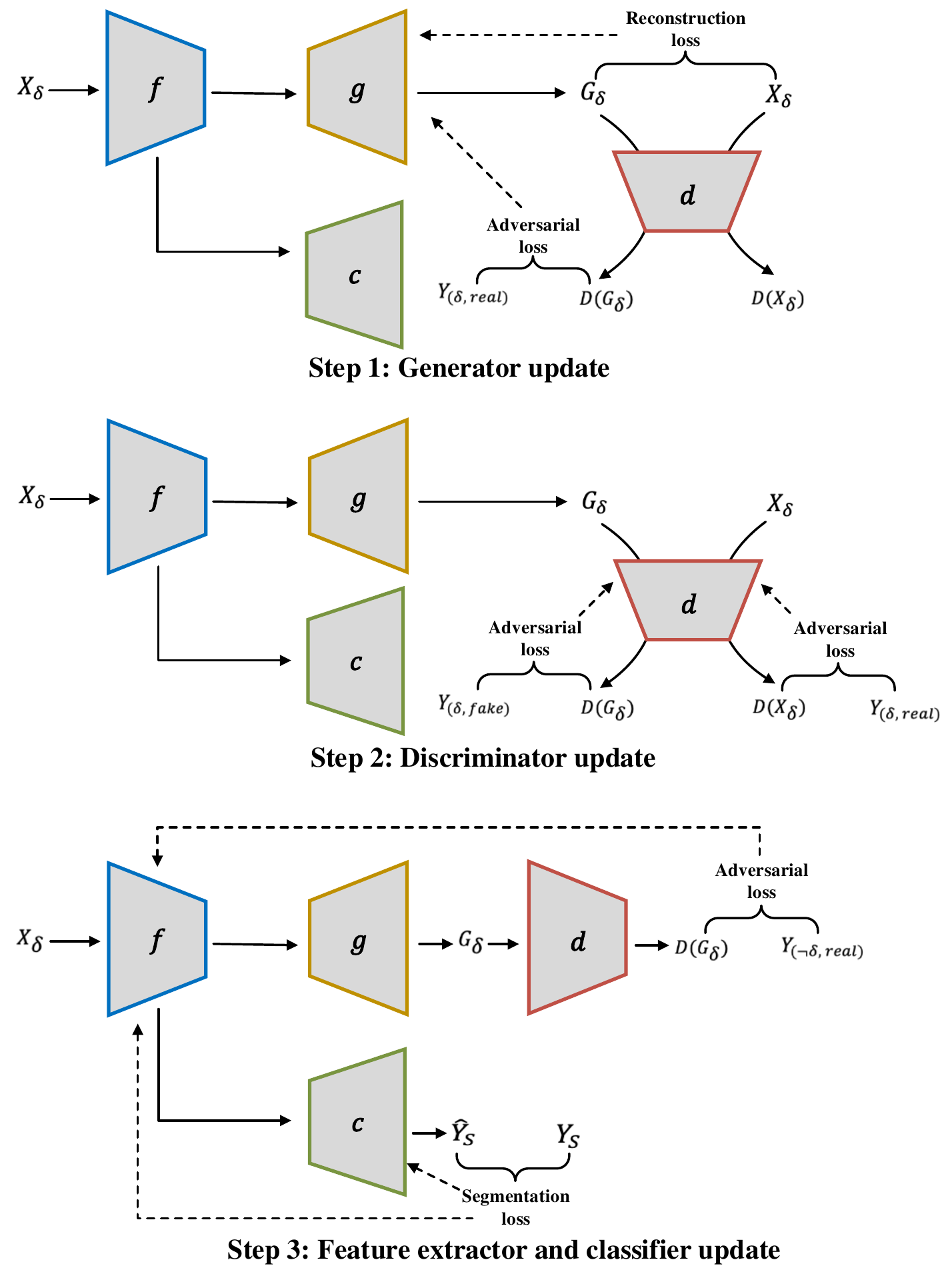}
\caption{Training process for the proposed network. The solid lines indicate the data flow, and the dashed lines indicate the gradient flow.}
\label{flowchart} 
\end{figure}

\subsection{Training and testing process.}
Our goal is to train a segmentation network that produces a competitive performance on real COVID-19 CT images even if no annotations are provided. We use the annotated synthetic images as the source and unlabeled real COVID-19 CT images as the target to jointly train the network, and update the parameters using segmentation loss, adversarial loss, and reconstruction loss. The segmentation loss is defined over adequately annotated source domain images, allowing the network to develop pixel-level discriminative ability. The adversarial loss can be divided into within-domain loss and cross-domain loss. The latter is used to guide the update of the feature extractor, thus allowing the feature extractor to identify the necessary features that should be extracted from the target domain. The reconstruction loss is utilized to ensure the fidelity of the reconstructions.

\noindent \textbf{Generator update.} The generator takes the learned features $F_S$ and $F_T$ as input, and reconstruct $X_S$ and $X_T$ as $G_S$ and $G_T$ conditioned on these feature maps. Intuitively, if the reconstruction is sufficiently accurate, there should be a low $L1$ loss between the reconstruction $G_\delta$ and input $X_\delta$. We also optimize the generator using adversarial loss, which forces the discriminator to classify $G_S$ and $G_T$ as real, thus fooling the discriminator. The object function of the generator can be represented by equation (\ref{eq4}),

\begin{equation}
\begin{aligned}
L_{G}=&\underbrace{\sum_\delta\sum_j{\left\|G_\delta-X_\delta \right\|}_1}_{reconstruction\;loss} -& \\
&\underbrace{\sum_\delta\sum_i{Y_{(\delta,real)}}logD(G_\delta)}_{within\-/domain\;adv\;loss}, \delta\in{(S,T)}&
\end{aligned}
\label{eq4}
\end{equation}

\noindent where index $i$ indicates the pixel location in the output probability map of discriminator and label map, index $j$ indicates the pixel location in the input and reconstruction.

\noindent \textbf{Discriminator update.} Given $X_S$, $X_T$, $G_S$, or $G_T$, the patch discriminator produces a 4-D probability map for each input. We calculate the adversarial loss using the cross-entropy loss between the output probability map and the label map $Y_{(\delta,\gamma)}, \delta\in(S,T), \gamma\in(real,fake)$. Therefore, the optimization process for the discriminator is as follows,

\begin{equation}
\begin{aligned}
L_{D}=&\underbrace{-\sum_\delta\sum_i{Y_{(\delta,real)}}logD(X_\delta)}_{within\-/domain\;adv\;loss} -& \\ 
&\underbrace{\sum_\delta\sum_i{Y_{(\delta,fake)}}logD(G_\delta)}_{within\-/domain\;adv\;loss}, \delta\in{(S,T)} &
\end{aligned}
\label{eq5}
\end{equation}

\noindent where $Y_{(\delta,\gamma)}$ is the $64\times64$ label map, in which each value corresponds to the label of each patch, indicating whether each patch of the input image belongs to the category of source-real, source-fake, target-real, or target-fake.

\noindent \textbf{Feature extractor and classifier update.} The updating of the feature extractor and classifier is a crucial process in our network for domain adaptation. We optimize these two components with the following combination of loss terms,

\begin{equation}
\begin{aligned}
L_{F}=&\underbrace{-\sum_c\sum_{j}Y_Slog(\hat{Y}_S)}_{segmentation\;loss} -& \\ &\underbrace{\alpha\sum_\delta\sum_i{Y_{(\neg\delta,real)}}logD(G_\delta)}_{cross\-/domain\;adv\;loss}, \delta\in{(S,T)}&
\end{aligned}
\label{eq6}
\end{equation}

\noindent where the first term is the segmentation loss. This is the pixel-wise cross-entropy loss calculated between the segmentation map and the annotation of the source domain. Index $c$ is the number of the categories in the segmentation results (four categories in our work: background, consolidation, ground-glass opacity, and the lung). 

Directly minimizing the segmentation loss in equation (\ref{eq6}) leads to good segmentation performance with the source images, but when tested on the target images, the performance will be significantly lower due to the domain shift. To overcome this problem, we introduce cross-domain adversarial loss to our network. Please note that, unlike the updating process for the generator and discriminator shown in equations (\ref{eq4}) and (\ref{eq5}), where the adversarial loss is calculated within the source or target domain. Here, the adversarial loss is cross-domain, and the gradients of the cross-domain adversarial loss can lead to a reversed domain classification. We utilize these gradients to update the feature extractor. To be more specific, the cross-domain adversarial losses are used to ensure that, when target features are passed to the generator, source-like images can be reconstructed, when source features are passed to the generator, target-like images can be reconstructed. Through this constraint domain alignment, the learned features from the two domains will exhibit a similar distribution, thus enabling the feature extractor to learn the common representations of the two domains.

In Figure \ref{flowchart}, we illustrate the training process for each module in the network with the direction of the data flow and gradient flow. For each iteration, the randomly sampled ${X_\delta,\delta\in{(S,T)}}$ are sent to the network. The generator, discriminator, feature extractor, and classifier are then iteratively updated in turn. Note that, unlike the updating of the generator and discriminator, the adversarial loss used to update the feature extractor and classifier is cross-domain. Except for the segmentation loss, all the other losses are calculated in the source domain and target domain. During the testing process, we only use the trained feature extractor and classifier. The network takes the real CT images of COVID-19 cases as input and generates the predicted segmentation map.

\section{Experiments}
\label{Experiments}
\subsection{Experimental settings}
\noindent \textbf{Dataset.} The source data comes from our previous work \cite{jiang2020covid}, which was designed to generate high-quality and realistic COVID-19 lung CT images for use in deep learning based medical imaging tasks. The dataset contains 10,200 synthetic 2D CT images with corresponding pixel-wise annotations. There are four categories in the annotation map: ground-glass opacity, consolidation, the lung, and background. The first two are the most common characteristics used for COVID-19 diagnosis in lung CT imaging. The target data is taken from the COVID-19 CT segmentation dataset \cite{covid19_dataset} collected by the Italian Society of Medical and International Radiology. It contains 9 CT volumes from confirmed COVID-19 patients, and each volume contains $\sim$200 slices. We reformated all 3D volumes into 2D slices with a size of $512\times512$. Small rotations, shearings, gamma transforms, and intensity normalizations are used for data augmentation, and there are a total of 12,000 slices after pre-processing. We employ $70\%$ slices as the unlabeled target data for training, while the remaining $30\%$ slices are used for testing segmentation performance. We follow the patient-level split rule when we separate the target data into training set and test set.

\noindent \textbf{Implementation details.} We use PyTorch \cite{paszke2017automatic} for implementation. Our network is trained with 100K iterations using Adam optimizer \cite{kingma2014adam}. The hyperparameters are set at $\alpha=0.1$ and $l_r=1.0\times10^{-5}$. The batch size is 1, and for every 10K iterations, the $l_r$ is reduced by ${20\%}$. The network training is accelerated with an NVIDIA RTX 2080Ti and an Intel(R) Core i7-9700K CPU.

\noindent \textbf{Evaluation metrics.}
For quantitative evaluation, we adopted the three most commonly used evaluation metrics in medical imaging analysis: the dice similarity coefficient (Dice), sensitivity (Sen), and specificity (Spe) \cite{fenster2006evaluation,milletari2016v}. The dice similarity coefficient is an overlap index that indicates the similarity between the prediction and the ground truth. Sensitivity and specificity are two statistical metrics for the performance of binary medical image segmentation tasks. The former measures the percentage of actual positive pixels correctly predicted to be positive, while the latter measures the proportion of actual negative pixels correctly predicted to be negative. These metrics are defined as follows:

\begin{equation}
\begin{aligned}
Dice = \frac{2\times{T\!P}}{2\times{T\!P}+{F\!P}+F\!N}\qquad
\end{aligned}
\label{eq7}
\end{equation}

\begin{equation}
\begin{aligned}
Sen = \frac{T\!P}{T\!P+F\!N}\qquad
\end{aligned}
\label{eq8}
\end{equation}

\begin{equation}
\begin{aligned}
Spe = \frac{T\!N}{T\!N+F\!P}\qquad
\end{aligned}
\label{eq9}
\end{equation}

\noindent where $T\!P$, $F\!P$, $T\!N$, and $F\!N$ represent the pixel number of true positive, false positive, true negative, and false negative in the prediction respectively.

\begin{table*}
\caption{Quantitative results for the two-class segmentation of COVID-19 CT images. Infection considers both ground-glass opacity and consolidation. \\(The highest evaluation score is marked in bold. $\uparrow$ indicates that a higher number is better.)}
\label{tab:two-class}       
\setlength{\tabcolsep}{2.5mm}{
\begin{tabular}{lllllll}
\hline
\multirow{2}*{Methods} & \multicolumn{3}{c}{Ground-glass opacity} & \multicolumn{3}{c}{Consolidation}  \\ \cline{2-7}
~ & Dice (\%) $\uparrow$ & Sen (\%) $\uparrow$ & Spe (\%) $\uparrow$ & Dice (\%) $\uparrow$ & Sen (\%) $\uparrow$ & Spe (\%) $\uparrow$ \\
\hline
Source-only & 80.60$\pm$0.48 & 78.86$\pm$0.52 & 99.60$\pm$0.01 & 61.75$\pm$0.50 & 66.73$\pm$0.51 & 99.83$\pm$0.01 \\
Self-ensembling \cite{perone2019unsupervised} & 82.43$\pm$0.36 & 80.18$\pm$0.47 & 99.53$\pm$0.01 & 65.16$\pm$0.78 & 66.58$\pm$1.26 & 99.26$\pm$0.01 \\
SSL \cite{ouyang2020self} & 78.34$\pm$0.87 & 71.37$\pm$0.53 & 99.47$\pm$0.01 & 73.83$\pm$0.91 & \textbf{81.30$\pm$0.82} & 99.44$\pm$0.01 \\
Ours & \textbf{85.34$\pm$0.36} & \textbf{82.13$\pm$0.41} & \textbf{99.87$\pm$0.01} & \textbf{74.67$\pm$0.57} & 68.69$\pm$0.39 & \textbf{99.97$\pm$0.01} \\
Target-only & 88.73$\pm$0.98 & 87.55$\pm$1.34 & 99.84$\pm$0.02 & 84.58$\pm$0.80 & 84.71$\pm$0.94 & 99.94$\pm$0.01 \\
\end{tabular}}

\setlength{\tabcolsep}{2.5mm}{
\begin{tabular}{lllllll}
\hline
\multirow{2}*{} & \multicolumn{3}{c}{Infection} & \multicolumn{3}{c}{Lung}  \\ \cline{2-7}
~ & Dice (\%) $\uparrow$ & Sen (\%) $\uparrow$ & Spe (\%) $\uparrow$ & Dice (\%) $\uparrow$ & Sen (\%) $\uparrow$ & Spe (\%) $\uparrow$ \\
\hline
Source-only & 78.82$\pm$0.61 & 70.99$\pm$0.86 & 99.80$\pm$0.01 & 89.60$\pm$0.62 & 92.38$\pm$0.23 & 97.89$\pm$0.15 \\
Self-ensembling \cite{perone2019unsupervised} & 80.43$\pm$0.47 & 80.74$\pm$0.51 & 99.63$\pm$0.01 & 93.53$\pm$0.29 & 90.47$\pm$0.10 & 99.61$\pm$0.01 \\
SSL \cite{ouyang2020self} & 79.15$\pm$0.51 & 78.77$\pm$0.50 & \textbf{99.81$\pm$0.01} & 94.59$\pm$0.19 & \textbf{93.47$\pm$0.13} & 97.60$\pm$0.01 \\
Ours & \textbf{86.54$\pm$0.39} & \textbf{85.54$\pm$0.43} & 99.80$\pm$0.01 & \textbf{95.75$\pm$0.25} & 93.11$\pm$0.26 & \textbf{99.74$\pm$0.01} \\
Target-only & 91.50$\pm$0.43 & 92.56$\pm$0.52 & 99.81$\pm$0.01 & 97.62$\pm$0.15 & 97.38$\pm$0.16 & 99.69$\pm$0.02 \\
\hline \\
\end{tabular}}
\end{table*}

\begin{table*}
\caption{Quantitative results for multi-class segmentation of COVID-19 CT images. Infection considers both ground-glass opacity and consolidation. \\(The highest evaluation score is marked in bold. $\uparrow$ means a higher number is better.)}
\label{tab:multi-class}       
\setlength{\tabcolsep}{3mm}{
\begin{tabular}{lllllll}
\hline
\multirow{2}*{Methods} & \multicolumn{3}{c}{Ground-glass opacity} & \multicolumn{3}{c}{Consolidation}  \\ \cline{2-7}
~ & Dice (\%) $\uparrow$ & Sen (\%) $\uparrow$ & Spe (\%) $\uparrow$ & Dice (\%) $\uparrow$ & Sen (\%) $\uparrow$ & Spe (\%) $\uparrow$ \\
\hline
Source-only & 79.16$\pm$0.56 & 73.65$\pm$0.41 & 99.81$\pm$0.01 & 61.42$\pm$0.45 & 57.54$\pm$0.67 & 99.82$\pm$0.01 \\
MinEnt \cite{vu2019advent} & 79.72$\pm$0.42 & 71.83$\pm$0.48 & 99.87$\pm$0.01 & \textbf{75.33$\pm$0.41} & 67.23$\pm$0.68 & \textbf{99.97$\pm$0.01} \\
AdvEnt \cite{vu2019advent} & 81.99$\pm$0.38 & 76.68$\pm$0.45 & 99.83$\pm$0.01 & 64.07$\pm$0.74 & 54.18$\pm$0.98 & 99.95$\pm$0.01 \\
IntraDA \cite{pan2020unsupervised} & 79.30$\pm$0.34 & 69.17$\pm$0.35 & \textbf{99.88$\pm$0.01} & 62.33$\pm$0.88 & 57.80$\pm$1.00 & \textbf{99.97$\pm$0.01} \\
Ours & \textbf{86.31$\pm$0.27} & \textbf{85.37$\pm$0.26} & 99.81$\pm$0.01 & 74.55$\pm$0.30 & \textbf{67.44$\pm$0.32} & {99.95$\pm$0.01} \\
Target-only & 87.54$\pm$0.27 & 86.83$\pm$0.34 & 99.82$\pm$0.01 & 84.88$\pm$0.42 & 82.79$\pm$0.62 & 99.96$\pm$0.01 \\
\end{tabular}}

\setlength{\tabcolsep}{3mm}{
\begin{tabular}{lllllll}
\hline
\multirow{2}*{} & \multicolumn{3}{c}{Infection} & \multicolumn{3}{c}{Lung}  \\ \cline{2-7}
~ & Dice (\%) $\uparrow$ & Sen (\%) $\uparrow$ & Spe (\%) $\uparrow$ & Dice (\%) $\uparrow$ & Sen (\%) $\uparrow$ & Spe (\%) $\uparrow$ \\
\hline
Source-only & 76.98$\pm$0.30 & 70.92$\pm$0.47 & 99.66$\pm$0.01 & 88.54$\pm$0.32 & 93.47$\pm$0.16 & 97.41$\pm$0.08 \\
MinEnt \cite{vu2019advent} & 80.91$\pm$0.27 & 72.61$\pm$0.30 & 99.86$\pm$0.01 & 95.55$\pm$0.01 & \textbf{95.62$\pm$0.01} & 99.33$\pm$0.01 \\
AdvEnt \cite{vu2019advent} & 81.12$\pm$0.28 & 74.55$\pm$0.35 & 99.82$\pm$0.01 & 95.69$\pm$0.06 & 95.41$\pm$0.05 & 99.41$\pm$0.01 \\
IntraDA \cite{pan2020unsupervised} & 77.34$\pm$0.32 & 67.76$\pm$0.43 & \textbf{99.89$\pm$0.01} & 95.27$\pm$0.07 & 95.01$\pm$0.06 & 99.35$\pm$0.01 \\
Ours & \textbf{86.15$\pm$0.29} & \textbf{84.29$\pm$0.31} & 99.81$\pm$0.01 & \textbf{96.13$\pm$0.07} & 94.61$\pm$0.09 & \textbf{99.67$\pm$0.01} \\
Target-only & 89.55$\pm$0.35 & 88.57$\pm$0.29 & 99.82$\pm$0.01 & 97.12$\pm$0.13 & 97.04$\pm$0.18 & 99.59$\pm$0.01 \\
\hline
\end{tabular}}
\end{table*}

\begin{figure*}[t]
\centering
\includegraphics[width=15cm, height=8.7cm]{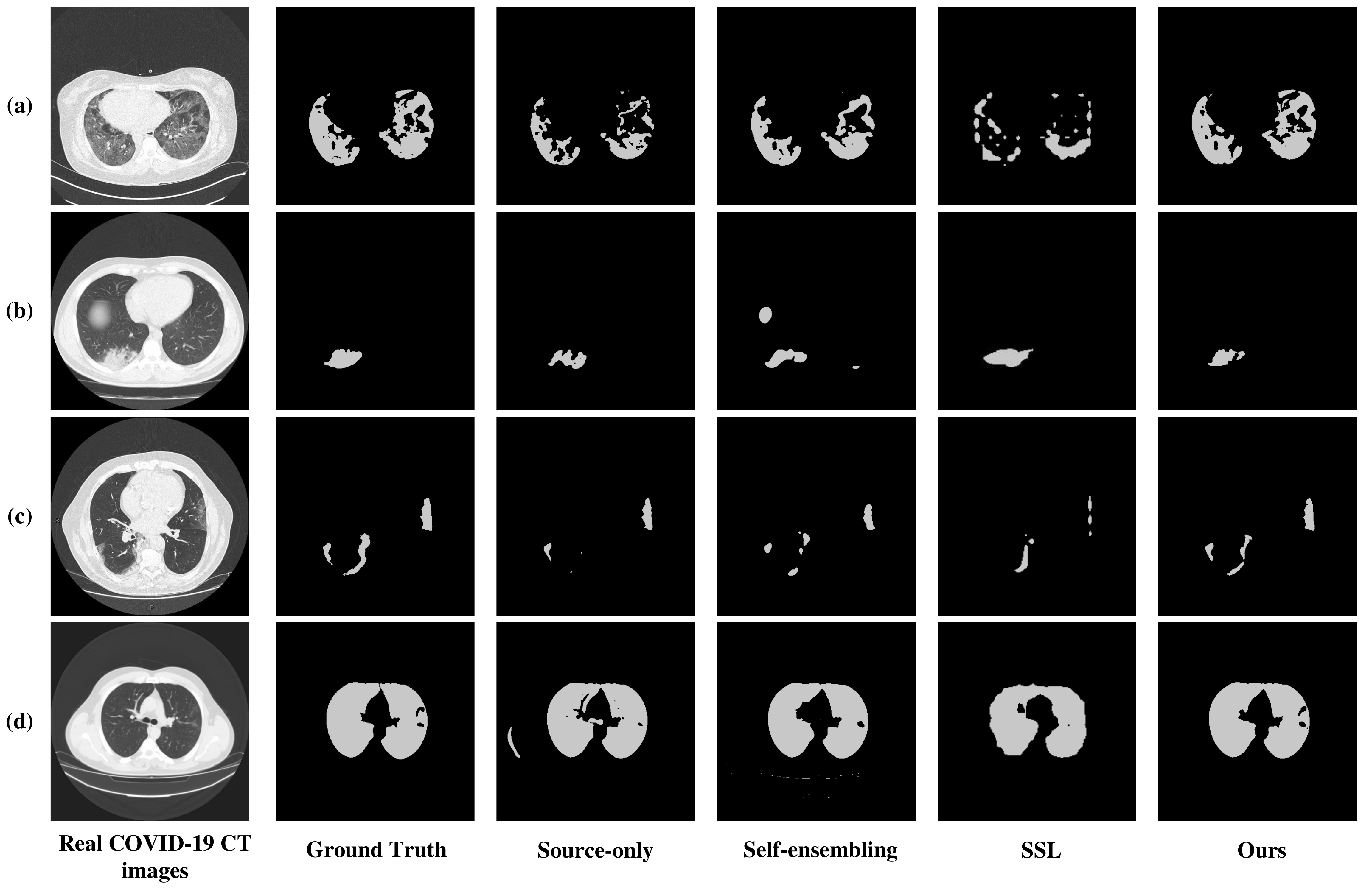}
\caption{Qualitative results for two-class segmentation task. Columns 1 and 2 present the input real COVID-19 CT images and corresponding ground truth, while Column 3 to 6 are segmentation results of Source-only, Self-ensembling\cite{perone2019unsupervised}, SSL\cite{ouyang2020self}, and our proposed method. The first to last rows are the results when taking ground-glass opacity (a), consolidation (b), infection (c) and the lung (d) as the segmentation object, respectively.}
\label{fig:two-class}       
\end{figure*}

\begin{figure*}[!h]
\centering
\includegraphics[width=17cm, height=11cm]{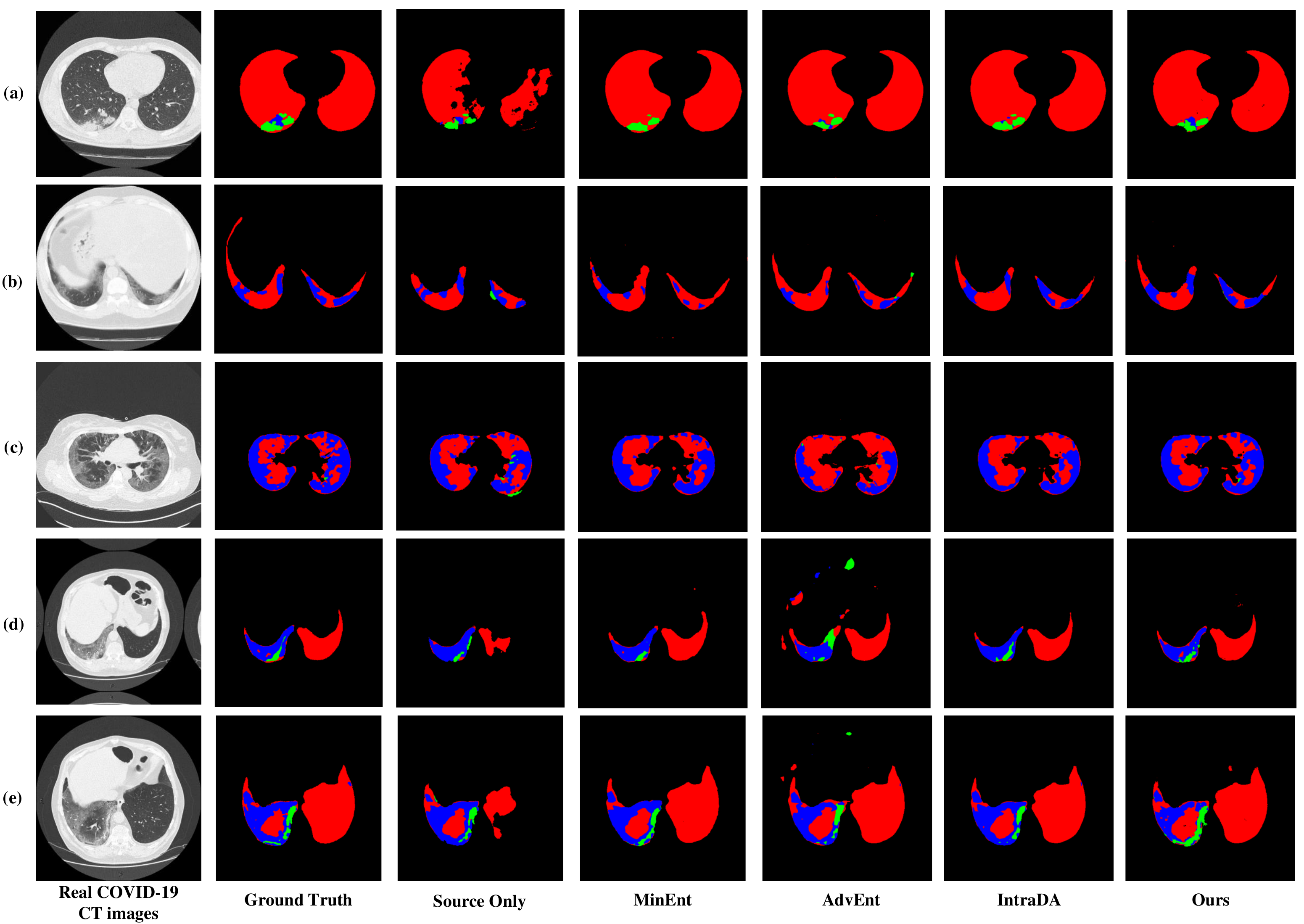}
\caption{Qualitative results for multi-class segmentation task. Columns 1 and 2 show the input real COVID-19 CT images and corresponding ground truth, in which the ground-glass opacity is marked in blue, consolidation is marked in green, and the lung is marked in red. Columns 3 to 7 are the segmentation results for the Source-only, MinEnt\cite{vu2019advent}, AdvEnt\cite{vu2019advent}, IntraDA\cite{pan2020unsupervised}, and our proposed method, respectively.}
\label{fig:multi-class}       
\end{figure*}

\subsection{Quantitative results}
\label{Quantitative results}
\noindent \textbf{Evaluation on two-class segmentation task.} In this section, we compare the segmentation performance of our proposal with two state-of-the-art unsupervised medical image segmentation methods: Self-ensembling \cite{perone2019unsupervised} and SSL \cite{ouyang2020self}. Because these methods are designed for two-class segmentation, we train our proposed approach as a two-class segmentation network, e.g., by taking the ground-glass opacity as the object to segment and other classes as the background.

Table \ref{tab:two-class} presents the experimental results when taking each category as the segmentation object. The results are reported as the mean $\pm$ error interval (calculated based on 95\% confidence interval). Our proposed method outperforms the reported methods across most metrics. Compared with the second-best method Self-ensembling \cite{perone2019unsupervised}, the proposed method produces a 6.11\% improvement in the dice similarity score for infection. Different with other compared methods, which utilize consistency loss to minimize the discrepancy between predictions in the source and target domain or employ superpixel-based pseudo-labels for supervision, our proposed approach attempts to learn the more discriminative feature representations when dealing with the challenging medical segmentation task.

\noindent \textbf{Evaluation on multi-class segmentation task.} As an assistant diagnostic tool, our model is expected to provide more detailed information about the infected areas and the lung. Therefore, we extend our method to a multi-class segmentation task, and compare it with state-of-the-art domain adaptation based segmentation methods MinEnt \cite{vu2019advent}, AdvEnt \cite{vu2019advent}, and IntraD \cite{pan2020unsupervised}. It should be noted that the metrics for each category are calculated by taking the other categories as background. More speciﬁcally, even though the network is trained for a multi-class segmentation task, we employ two classes (the object and background) when calculating the metrics.

Table \ref{tab:multi-class} shows the quantitative results on real CT images from COVID-19 cases. Our proposed approach outperforms the compared methods across most metrics. Compared with the second-best method AdvEnt \cite{vu2019advent}, our proposal produces a $5.03\%$ improvement in the dice similarity score for infection. When excluding the domain adaptation module of our network, and only using the base feature extractor and classifier trained with the source data (source-only), we observe a significant drop in performance (Dice: $86.15\%\rightarrow76.98\%$ for infection), clearly illustrating the effectiveness of our domain adaptation strategy, which employs adversarial training to learn the true features of the infection from real COVID-19 CT images. It can also be observed that, even without access to the ground truth for the real CT images, our proposed method achieves results that are comparable to the target-only method trained with the target data in a supervised manner. Moreover, our proposal achieves the highest performance in lung segmentation. This proves that the proposed method is also suitable for large-area tissues or organ segmentation.

\subsection{Qualitative results}
Figure \ref{fig:two-class} shows the qualitative results for two-class segmentation of real COVID-19 CT images. We train our method as a two-class segmenter for ground-glass opacity, consolidation, infection, and the lung respectively. It is obvious that the proposed domain adaptation based segmentation network can learn the discriminative features by employing the adversarial training, so as to accurately segment the object areas. The Self-ensembling \cite{perone2019unsupervised} can handle the large object segmentation such as (a) and (d), but demonstrates a poor performance for the relatively small consolidation shown in (b). SSL \cite{ouyang2020self} relies on the superpixel-based pseudo labels for supervision during training, so it fails to capture the details of ground-glass opacity in (a).

Figure \ref{fig:multi-class} displays the qualitative results for multi-class segmentation of real COVID-19 CT cases. It is obvious that there are a large number of mis-segmented areas in the visualization results of the source-only (baseline) model. This is mainly due to the differences in the texture and intensity between the synthetic data and the real COVID-19 CT images. We observe a significant improvement in performance when introducing cross-domain adversarial learning in our proposed approach, which confirms the importance of adversarial training based domain adaptation. MinEnt \cite{vu2019advent} attempts to minimize the entropy value of the model output directly to overcome the domain gap. However, compared with our proposed method, MinEnt fails to capture the fine-grained details of the infection in (a) and (b). AdvEnt \cite{vu2019advent} conducts adversarial training on entropy map and it is quite sensitive to the influence of irrelevant areas, for example, there is obvious noise in the results (d) and (e). IntraDA \cite{pan2020unsupervised} relies on the pseudo labels for training, thus it fails to separate the ground-glass opacity and consolidation in (a).

Our domain adaptation based segmentation network outperforms the baseline method and other state-of-the-art methods. It produces a performance that is close to the ground truth with fewer mis-segmented infection areas, especially for consolidation, which is relatively small and challenging to segment. The success of the proposed method is attributed to our adversarial training scheme, through which our network can learn the true features of target data under the constraint of cross-domain adversarial loss. This scheme allows our network to more clearly distinguish the real features of ground-glass opacity and consolidation even without access to ground truth annotations of the target data. In addition, our proposed method also performs best in terms of lung segmentation, which proves that our method can be generalized. That is, it can be used not only for COVID-19 infection segmentation, but also for other organs.

\begin{table*}\centering
\caption{Ablation study of different feature map selection strategies. \\(The highest evaluation score is marked in bold. $\uparrow$ means a higher number is better. GGO: ground-glass opacity)}
 \label{tab:ablation-1}
\setlength{\tabcolsep}{2mm}{
\begin{tabular}{ccccccccc}
\hline
\multirow{2}*{1} & \multirow{2}*{2} & \multirow{2}*{3} & \multirow{2}*{4} & \multirow{2}*{5} & \multicolumn{4}{c}{Dice (\%) $\uparrow$} \\ \cline{6-9}
~ & ~ & ~ & ~ & ~ & GGO & Consolidation & Infection & Lung \\
\hline
$\checkmark$ & $\checkmark$ & $\checkmark$ & $\times$ & $\times$ & 83.43$\pm$0.40 & 67.35$\pm$0.36 & 83.30$\pm$0.23 & 95.69$\pm$0.07 \\
$\times$ & $\checkmark$ & $\checkmark$ & $\checkmark$ & $\times$ & 85.36$\pm$0.28 & \textbf{74.83$\pm$0.40} & 84.24$\pm$0.16 & 96.11$\pm$0.08 \\
  $\times$ & $\times$ & $\checkmark$ & $\checkmark$ & $\checkmark$ & \textbf{86.31$\pm$0.27} & 74.55$\pm$0.30 & \textbf{86.15$\pm$0.29} & \textbf{96.13$\pm$0.07} \\
\hline
\end{tabular}}
\end{table*}

\begin{table*}\centering
\caption{Ablation study of different components in the proposed network. \\(The highest evaluation score is marked in bold. $\uparrow$ means a higher number is better. GGO: ground-glass opacity)}
 \label{tab:ablation-2}
\setlength{\tabcolsep}{2mm}{
\begin{tabular}{lcccc}
\hline
\multirow{2}*{Network configuration} & \multicolumn{4}{c}{Dice (\%) $\uparrow$} \\ \cline{2-5}
~ & GGO & Consolidation & Infection & Lung \\
\hline
w/o adversarial training & 79.16$\pm$0.56 & 61.42$\pm$0.45 & 79.98$\pm$0.30 & 88.54$\pm$0.32 \\
w/o skip connections & 78.56$\pm$0.48 & 61.71$\pm$0.63 & 78.81$\pm$0.24 & 94.11$\pm$0.01 \\
Feature space training & 81.87$\pm$0.33 & 52.61$\pm$0.43 & 79.52$\pm$0.41 & 92.33$\pm$0.22 \\
\textbf{Ours} & \textbf{86.31$\pm$0.27} & \textbf{74.55$\pm$0.30} & \textbf{86.15$\pm$0.29} & \textbf{96.13$\pm$0.07} \\
\hline
\end{tabular}}
\end{table*}

\subsection{Ablation study}
\label{Ablation study}
In order to assess the important settings and components of our method, we conduct ablation experiments following the multi-class experimental settings in sub-section \ref{Quantitative results}. The evaluation criterion is the dice similarity coefficient.

\noindent \textbf{Comparison of different feature map selection strategies.} As described in sub-section \ref{Network structure}, the generator takes three different feature maps from the feature extractor as the input and maps them to the image space. Then, the output of the generator is used to calculate the adversarial loss, which is crucial for domain-invariant feature learning. Therefore, the selection strategy for the feature maps will affect the segmentation performance. Because there are four down-sampling operations in our feature extractor, we have a total of five sizes including the original image (1:$512\times512$, 2:$256\times256$, 3:$128\times128$, 4:$64\times64$, 5:$32\times32$). We conduct a series of experiments using different combinations. From Table \ref{tab:ablation-1}, it can be observed that our network achieves the highest performance when the generator takes high-level feature maps as the input. This proves that the high-level semantic information is more helpful for domain adaptation than the rich details in the low-level features maps. We adopt this setting for our network.

\noindent \textbf{Effect of different components in our network.} Table \ref{tab:ablation-2} shows how the different components of our network influence the segmentation performance. The bold line corresponds to our proposed method, while the other methods differ from our proposed approach in the following respects. (1) w/o adversarial training: the source-only baseline corresponds to $\alpha$=0. Here, the domain adaptation module is excluded and only the feature extractor and classifier are used. (2) w/o skip connections: the skip connections between the feature extractor and classifier are removed, which are essential for preserving the fine-grained details in the segmentation. (3) feature space training: the domain adaptation module is removed and pixel-level adversarial loss for the feature maps is calculated, which is then used to update the feature extractor and classifier. The experimental results show that the domain adaptation module is critical to ensuring the excellent performance of our network. In addition, compared with feature space training, calculating the adversarial loss on image space is more efficient.

\section{Conclusion}
\label{Conclusion}
In this paper, we proposed a novel unsupervised domain adaptation based method for COVID-19 infection segmentation in CT images. We considered a challenging situation in which abundant synthetic annotated medical images are available, but no annotations are available for real COVID-19 lung CT images. We introduced unsupervised adversarial training to our network to correlate the features between real COVID-19 CT images and synthetic images. The cross-domain adversarial loss enforces the features learned by feature extractor from the two domains closer, thus the network can learn the common representations of two domains and retain the diagnostic information (i.e., the features of COVID-19 infection). Experimental results on CT images of COVID-19 cases demonstrated that our proposal outperforms baseline and state-of-the-art approaches. We also demonstrated the effectiveness of our network in lung segmentation. Our proposed method has great potential for use in diagnosing COVID-19 by quantifying the infected areas of the lung.

\bibliographystyle{unsrt}
\bibliography{ref}
\end{document}